\documentclass[ALICE,manyauthors]{cernphprep}
\usepackage[comma,square,numbers,sort&compress]{natbib}
\usepackage{comment}

\usepackage{lineno}
\usepackage{xspace}
\usepackage{hyperref}
\usepackage{multirow}
\usepackage{color}
\usepackage[T1]{fontenc}

\begin{document}
%

\newcommand{\pp}           {pp\xspace}
\newcommand{\pplong}       {proton-proton\xspace}
\newcommand{\ppbar}        {\mbox{$\mathrm {p\overline{p}}$}\xspace}
\newcommand{\ee}           {\mbox{$\mathrm {e^{+}e^{-}}$}\xspace}
\newcommand{\ep}           {\mbox{$\mathrm {e^{-}p}$}\xspace}
\newcommand{\XeXe}         {\mbox{Xe--Xe}\xspace}
\newcommand{\PbPb}         {\mbox{Pb--Pb}\xspace}
\newcommand{\pA}           {\mbox{pA}\xspace}
\newcommand{\pPb}          {\mbox{p--Pb}\xspace}
\newcommand{\AuAu}         {\mbox{Au--Au}\xspace}
\newcommand{\dAu}          {\mbox{d--Au}\xspace}
\let\eepm=\ee
\newcommand{\hi}           {heavy-ion\xspace}
\newcommand{\Hi}           {Heavy-ion\xspace}

\newcommand{\s}            {\ensuremath{\sqrt{s}}\xspace}
\newcommand{\snn}          {\ensuremath{\sqrt{s_{\mathrm{NN}}}}\xspace}
\newcommand{\pt}           {\ensuremath{p_{\rm T}}\xspace}
\newcommand{\kt}           {\ensuremath{k_{\rm T}}\xspace}
\newcommand{\meanpt}       {$\langle p_{\mathrm{T}}\rangle$\xspace}
\newcommand{\ycms}         {\ensuremath{y_{\rm CMS}}\xspace}
\newcommand{\ylab}         {\ensuremath{y_{\rm lab}}\xspace}
\newcommand{\etarange}[1]  {\mbox{$\left | \eta \right |~<~#1$}\xspace}
\newcommand{\yrange}[1]    {\mbox{$\left | y \right |~<~#1$}\xspace}
\newcommand{\dndy}         {\ensuremath{\mathrm{d}N_\mathrm{ch}/\mathrm{d}y}\xspace}
\newcommand{\dndeta}       {\ensuremath{\mathrm{d}N_\mathrm{ch}/\mathrm{d}\eta}\xspace}
\newcommand{\avdndeta}     {\ensuremath{\langle\dndeta\rangle}\xspace}
\newcommand{\avdndetaetarange}     {\ensuremath{\langle\dndeta\rangle_{|\eta| <0.5}}\xspace}
\newcommand{\dNdy}         {\ensuremath{\mathrm{d}N_\mathrm{ch}/\mathrm{d}y}\xspace}
\newcommand{\Npart}        {\ensuremath{N_\mathrm{part}}\xspace}
\newcommand{\Ncoll}        {\ensuremath{N_\mathrm{coll}}\xspace}
\newcommand{\dEdx}         {\ensuremath{\textrm{d}E/\textrm{d}x}\xspace}
\newcommand{\RpPb}         {\ensuremath{R_{\rm pPb}}\xspace}
\newcommand{\DeltaM}       {\ensuremath{\Delta M}\xspace}
\let\dNchdeta=\avdndeta
\newcommand{\ntrkl}        {\ensuremath{N_\mathrm{trkl}}\xspace}
\newcommand{\vzeromperc}   {\ensuremath{p_\mathrm{V0M}}\xspace}
\newcommand{\zvtx}        {\ensuremath{z_\mathrm{vtx}}\xspace}

\newcommand{\sqrts}        {\ensuremath{\sqrt{s}}\xspace}
\newcommand{\sqrtsNN}      {\ensuremath{\sqrt{s_{\mathrm{NN}}}}\xspace}
\newcommand{\nineH}        {$\sqrt{s}=0.9$~Te\kern-.1emV\xspace}
\newcommand{\seven}        {$\sqrt{s}=7$~Te\kern-.1emV\xspace}
\newcommand{\twoH}         {$\sqrt{s}=0.2$~Te\kern-.1emV\xspace}
\newcommand{\twosevensix}  {$\sqrt{s}=2.76$~Te\kern-.1emV\xspace}
\newcommand{\five}         {$\sqrt{s}=5.02$~Te\kern-.1emV\xspace}
\newcommand{\thirteen}     {$\sqrt{s}=13$~Te\kern-.1emV\xspace}
\newcommand{\twosevensixnn}{$\sqrt{s_{\mathrm{NN}}}=2.76$~Te\kern-.1emV\xspace}
\newcommand{\fivenn}       {$\sqrt{s_{\mathrm{NN}}}=5.02$~Te\kern-.1emV\xspace}
\newcommand{\twoHnn}       {$\sqrt{s_{\mathrm{NN}}}=200$~Ge\kern-.1emV\xspace}
\newcommand{\LT}           {L{\'e}vy-Tsallis\xspace}
\newcommand{\GeVc}         {Ge\kern-.1emV/$c$\xspace}
\newcommand{\MeVc}         {Me\kern-.1emV/$c$\xspace}
\newcommand{\TeV}          {Te\kern-.1emV\xspace}
\newcommand{\GeV}          {Ge\kern-.1emV\xspace}
\newcommand{\MeV}          {Me\kern-.1emV\xspace}
\newcommand{\GeVcc}        {Ge\kern-.2emV/$c^2$\xspace}
\newcommand{\MeVcc}        {Me\kern-.2emV/$c^2$\xspace}
\newcommand{\lumi}         {\ensuremath{\mathcal{L}}\xspace}
\newcommand{\degree}       {\ensuremath{^{\rm o}}\xspace}
\let\gevc=\GeVc
\let\mevc=\MeVc
\let\tev=\TeV
\let\gev=\GeV
\let\mev=\MeV
\let\gevcc=\GeVcc
\let\mevcc=\MeVcc
\let\GeVmom=\GeVc
\let\MeVmom=\MeVc
\let\GeVmass=\GeVcc
\let\MeVmass=\MeVcc

\newcommand{\ITS}          {\rm{ITS}\xspace}
\newcommand{\TOF}          {\rm{TOF}\xspace}
\newcommand{\ZDC}          {\rm{ZDC}\xspace}
\newcommand{\ZDCs}         {\rm{ZDCs}\xspace}
\newcommand{\ZNA}          {\rm{ZNA}\xspace}
\newcommand{\ZNC}          {\rm{ZNC}\xspace}
\newcommand{\SPD}          {\rm{SPD}\xspace}
\newcommand{\SDD}          {\rm{SDD}\xspace}
\newcommand{\SSD}          {\rm{SSD}\xspace}
\newcommand{\TPC}          {\rm{TPC}\xspace}
\newcommand{\TRD}          {\rm{TRD}\xspace}
\newcommand{\VZERO}        {\rm{V0}\xspace}
\newcommand{\VZEROA}       {\rm{V0A}\xspace}
\newcommand{\VZEROC}       {\rm{V0C}\xspace}

\newcommand{\pip}          {\ensuremath{\pi^{+}}\xspace}
\newcommand{\pim}          {\ensuremath{\pi^{-}}\xspace}
\newcommand{\kap}          {\ensuremath{\rm{K}^{+}}\xspace}
\newcommand{\kam}          {\ensuremath{\rm{K}^{-}}\xspace}
\newcommand{\pbar}         {\ensuremath{\rm\overline{p}}\xspace}
\newcommand{\kzero}        {\ensuremath{{\rm K}^{0}_{\rm{S}}}\xspace}
\newcommand{\lmb}          {\ensuremath{\Lambda}\xspace}
\newcommand{\almb}         {\ensuremath{\overline{\Lambda}}\xspace}
\newcommand{\Om}           {\ensuremath{\Omega^-}\xspace}
\newcommand{\Mo}           {\ensuremath{\overline{\Omega}^+}\xspace}
\newcommand{\X}            {\ensuremath{\Xi^-}\xspace}
\newcommand{\Ix}           {\ensuremath{\overline{\Xi}^+}\xspace}
\newcommand{\Xis}          {\ensuremath{\Xi^{\pm}}\xspace}
\newcommand{\Oms}          {\ensuremath{\Omega^{\pm}}\xspace}
\newcommand{\Vdecay} 	   {\ensuremath{\rm V^{0}}\xspace}

\newcommand{\Dzero}        {\ensuremath{\rm D^{0}}\xspace}
\newcommand{\Dplus}        {\ensuremath{\rm D^{+}}\xspace}
\newcommand{\Dstar}        {\ensuremath{\rm D^{*+}}\xspace}
\newcommand{\Ds}           {\ensuremath{\rm D^{+}_{\rm s}}\xspace}
\newcommand{\Lambdac}      {\ensuremath{\rm \Lambda_{\rm c}^{+}}\xspace}
\newcommand{\Lambdab}      {\ensuremath{\rm \Lambda_{\rm b}^{0}}\xspace}
\let\Lc=\Lambdac
\let\Lb=\Lambdab
\newcommand{\XicZeroPlus}            {\ensuremath{\rm \Xi_{c}^{0,+}}\xspace}
\newcommand{\XicPlus}                {\ensuremath{\rm \Xi_{c}^{+}}\xspace}
\newcommand{\Sigmac}                 {\ensuremath{\rm \Sigma_{c}}\xspace}
\newcommand{\SigmacZeroPlusPlus}     {\ensuremath{{\rm \Sigma_{c}^{0,++}}}\xspace}
\newcommand{\SigmacZero}             {\ensuremath{{\rm \Sigma_{c}^{0}}}\xspace}
\newcommand{\SigmacPlusPlus}         {\ensuremath{{\rm \Sigma_{c}^{++}}}\xspace}
\newcommand{\SigmacZeroPlusPlusPlus} {\ensuremath{\rm \Sigma_{c}^{0,+,++}}\xspace}

\newcommand{\DtoKpi}          {\ensuremath{\rm D^{0} \to K^{-}\pi^{+}}\xspace}
\newcommand{\Dstophip}        {\ensuremath{\rm D_{\rm s}^{+} \to \phi \pi^{+} \to K^{+}K^{-}\pi^{+}}\xspace}
\newcommand{\LctopKpi}        {\ensuremath{\rm \Lambda_{\rm c}^{+} \to pK^{-}\pi^{+}}\xspace}
\newcommand{\LctopKzeros}     {\ensuremath{\rm \Lambda_{\rm c}^{+} \to pK^{0}_{\rm S}}\xspace}
\newcommand{\LctopKzerosfull} {\ensuremath{\rm \Lambda_{\rm c}^{+} \to pK^{0}_{\rm S} \to p\pi^{+}\pi^{-}}\xspace}
\newcommand{\pKpi}            {\ensuremath{\rm pK^{-}\pi^{+}}\xspace}
\newcommand{\pKzeros}         {\ensuremath{\rm pK^{0}_{\rm S}}\xspace}
\let\LambdactopKpi=\LctopKpi
\let\LambdactopKzeros=\LctopKzeros
\let\LambdactopKzerosfull=\LctopKzerosfull

\newcommand{\DsDzero}         {\ensuremath{\Ds/\Dzero}\xspace}
\newcommand{\LcDzero}         {\ensuremath{\Lc/\Dzero}\xspace}
\let\LcD=\LcDzero
\newcommand{\inel}            {\ensuremath{\mathrm{INEL>0}}\xspace}
\let\INEL=\inel

\newcommand{\Hf}            {Heavy-flavour\xspace}
\newcommand{\hf}            {heavy-flavour\xspace}

\newcommand{\alphas}        {\ensuremath{\alpha_\mathrm{S}\xspace}}

\newcommand{\secletter}[1]    {\vspace{0.3cm}\textit{#1}}

\begin{titlepage}
\PHyear{2021}       
\PHnumber{245}      
\PHdate{23 November}  

\title{Observation of a multiplicity dependence in the \pt-differential charm baryon-to-meson ratios in proton--proton collisions at $\sqrt{s} = 13$~TeV}
\ShortTitle{Charm-hadron yield ratios versus multiplicity in pp at $\sqrt{s} = 13$~TeV}

\Collaboration{ALICE Collaboration\thanks{See Appendix~\ref{app:collab} for the list of collaboration members}}
\ShortAuthor{ALICE Collaboration} 

\begin{abstract}
The production of prompt \Dzero, \Ds, and \Lc hadrons, and their ratios, \DsDzero and \LcDzero, are measured in proton--proton collisions at \thirteen at midrapidity ($|y| <0.5$) with the ALICE detector at the LHC. The measurements are performed as a function of the charm-hadron transverse momentum (\pt) in intervals of charged-particle multiplicity, measured with two multiplicity estimators covering different pseudorapidity regions.
While the strange to non-strange \DsDzero ratio indicates no significant multiplicity dependence, the baryon-to-meson \pt-differential \LcDzero ratio shows a multiplicity-dependent enhancement, with a significance of 5.3$\sigma$ for $1<\pt < 12$~\gevc, comparing the highest multiplicity interval with respect to the lowest one. The measurements are compared with a theoretical model that explains the multiplicity dependence by a canonical treatment of quantum charges in the statistical hadronisation approach, and with predictions from event generators that implement colour reconnection mechanisms beyond the leading colour approximation to model the hadronisation process. 
The \LcDzero ratios as a function of \pt present a similar shape and magnitude as the $\Lambda/\kzero$ ratios in comparable multiplicity intervals, suggesting a potential common mechanism for light- and charm-hadron formation, with analogous multiplicity dependence. The \pt-integrated ratios, extrapolated down to $\pt=0$, do not show a significant dependence on multiplicity within the uncertainties.
\end{abstract}
\end{titlepage}

\setcounter{page}{2} 

\section{Introduction}
Heavy-flavour hadrons are produced in high-energy particle collisions through the hadronisation of the corresponding heavy-flavour quarks, which in turn typically originate from early hard scattering processes. The most common theoretical approach to describe this production is based on the quantum chromodynamics (QCD) factorisation theorem~\cite{Collins:1989gx}. In this framework, the production of hadrons containing charm or beauty quarks is calculated as a convolution of three independent terms: the parton distribution functions of the incoming protons, the cross sections of the partonic scatterings producing the heavy quarks, and the fragmentation functions that parametrise the non-perturbative evolution of a heavy quark into a given species of heavy-flavour hadron. Calculations based on the factorisation approach rely on the assumption that fragmentation functions, which are typically measured in electron--positron (\ee) or electron--proton (\ep) collisions~\cite{Braaten:1994bz}, are universal across all collision systems and energies. Systematic measurements of the relative production of heavy-flavour hadrons performed in different collision systems provide an excellent experimental benchmark to test this assumption.
 
Perturbative calculations at next-to-leading order, with next-to-leading-log resummation~\cite{Kniehl:2005mk,Kniehl:2012ti,Cacciari:1998it,Cacciari:2012ny}, can successfully describe the production cross section of strange and non-strange charm mesons and their ratios, as a function of transverse momentum (\pt) and rapidity in proton--proton (pp) collisions, over a wide range of centre-of-mass energies~\cite{Andronic:2015wma,Aaij:2015bpa,Khachatryan:2016csy,Aaij:2017qml,Acharya:2019mgn}. In contrast, these calculations, which are based on collinear factorisation and fragmentation functions tuned on \ee and \ep collision measurements, provide a poor description of heavy-flavour baryon production in hadronic collisions. Measurements of the \Lc production cross section in pp collisions at centre-of-mass energies of $\sqrts=7$, 5.02, and 13~TeV~\cite{Acharya:2017kfy,PhysRevC.104.054905,Sirunyan:2019fnc, Acharya:2021vpo} have shown a larger \pt-differential cross section in the measured \pt range, compared to QCD calculations~\cite{Kniehl:2005mk,Kniehl:2012ti,Kniehl:2020szu} as well as higher values for the \LcD ratio with respect to \ee collision data from LEP~\cite{Gladilin:2014tba}. Similarly, a \LcD ratio larger than expectations from \ee collisions was measured in p--Pb collisions at the LHC, both at midrapidity by ALICE~\cite{Acharya:2017kfy,PhysRevC.104.054905} and at forward rapidity by the LHCb experiment~\cite{LHCb:2018weo}.

In particular, the measurements in pp collisions at $\sqrts=5.02$ and 13~TeV provided the statistical precision to discriminate among different theoretical approaches. The measurements show, with good accuracy, a decrease of the \LcDzero ratio from about 0.6 in the interval $1<\pt<2$~\GeVc to about 0.3 for $8<\pt<12$~\GeVc. Calculations based on PYTHIA~8~\cite{Sjostrand:2014zea} with Monash tune~\cite{Skands:2014pea} and HERWIG~7~\cite{Bellm:2015jjp}, in which the charm fragmentation is tuned to \ee and \ep measurements, cannot describe the experimental results since they predict a \pt-independent \LcD ratio of about $0.1$. The Monash tune is based on the Lund string fragmentation model~\cite{Andersson:1997xwk, Andersson:1983ia}, where quarks and gluons, connected by colour strings, fragment into hadrons, and colour reconnection allows for partons created in the collision to interact via colour strings. A much better agreement is achieved by PYTHIA~8 calculations that include colour reconnection mechanisms beyond the leading-colour approximation~\cite{Christiansen:2015yqa} (CR-BLC in the following). These hadronisation mechanisms are implemented in addition to those included in the standard Monash tune. The CR-BLC calculations introduce new colour-reconnection topologies enhancing the contribution of ``junctions'' that fragment into baryons, thus providing an augmented baryon production. Calculations based on the statistical hadronisation model~\cite{He:2019tik} or calculations that include mechanisms of charm-hadron formation through coalescence of constituent quarks in the presence of a colour-deconfined state of matter~\cite{Minissale:2020bif}, also provide a satisfactory description of the \LcD ratio in pp collisions. This suggests the presence of modified or additional hadronisation mechanisms in small hadronic collision systems with respect to fragmentation in vacuum. Similar conclusions are drawn from recent measurements of higher-mass charm-baryon states, \XicZeroPlus and \SigmacZeroPlusPlus, in pp collisions at $\sqrts=5.02$, 7, and 13~\TeV~\cite{Acharya:2017lwf,Acharya:2021dsq,Acharya:2021vjp,Acharya:2021vpo}. The fragmentation fractions, i.e.~the probabilities for a charm quark to hadronise into a specific charm hadron, computed for the first time from  hadronic collision measurements at the LHC including the charm baryon states, are found to be different than those measured in \ee and \ep collisions. This observation confirms that the hadronisation of charm quarks into charm hadrons is not a universal process among different collision systems~\cite{Acharya:2021set}. 

The measurements of the \LcD and \DsDzero ratios also play an important role in the study of heavy-ion collisions, where a hot and dense quark--gluon plasma, characterised by the presence of free colour charges, is formed~\cite{Busza:2018rrf}. In heavy-ion collisions, measurements of baryon-to-meson ratios and of strange to non-strange hadron production ratios~\cite{Abelev:2013xaa,Adams:2006wk,Acharya:2018ckj,Adam:2019hpq,Sirunyan:2019fnc,ALICE:2021rxa, Acharya:2018hre,ALICE:2021kfc,STAR:2021tte} are sensitive to the mechanisms of hadronisation from the quark--gluon plasma~\cite{Fries:2008hs}. A first measurement of the \LcD ratio in Pb--Pb collisions, in the 80\% most central collisions, was performed at \fivenn~\cite{Acharya:2018ckj} by ALICE. The measurement is consistent with the hypothesis of an enhancement of the \LcD ratio with respect to pp collisions in the intermediate \pt region $6<\pt<12$~\GeVc, although the limited statistical precision does not yet allow for a firm conclusion to be drawn. 
The \LcD ratio in heavy-ion collisions was also measured by CMS, in Pb--Pb collisions at \fivenn for $10<\pt<20$~\gevc~\cite{Sirunyan:2019fnc}, and by STAR, in Au--Au collisions at \twoHnn for $2.5<\pt<8$~\gevc~\cite{Adam:2019hpq}. While the STAR result is significantly higher than PYTHIA~8 calculations with different tunes~\cite{Skands:2014pea, Christiansen:2015yqa}, the CMS ratio at higher \pt is consistent with the pp result.
A hint of enhancement of the \DsDzero ratio in central Pb--Pb collisions with respect to pp collisions was also observed at \fivenn in the intermediate \pt region $4<\pt<8$~\GeVc, as expected in the presence of a sizeable contribution of coalescence processes and increased strangeness production in the medium~\cite{Acharya:2018hre,ALICE:2021kfc}. 
A similar conclusion is drawn by STAR from the measured \DsDzero ratio in the 10\% most central Au--Au collisions at \twoHnn relative to PYTHIA simulation of pp collisions~\cite{STAR:2021tte}.
A measurement performed in high-multiplicity pp collisions could shed light on the possible presence of similar effects also in smaller collision systems with large particle densities. 

In this Letter, we present the first measurement of the production yields of prompt \Dzero, \Ds and \Lc (i.e.~produced in the hadronization of charm quarks or from the decay of excited open charm and charmonium states) as well as corresponding ratios, \DsDzero and \LcDzero, in pp collisions at \thirteen, as a function of the charged-particle pseudorapidity density \avdndeta. The aim of this study is to characterise the evolution of the aforementioned ratios from very low to moderate charged-particle density and provide new experimental constraints on the nature of these modifications in pp collisions. The study was performed considering events selected according to the charged-particle density at mid- and forward rapidities, in order to investigate the effects of possible biases originating from the determination of the multiplicity in the same pseudorapidity region in which charm hadrons are reconstructed. 
Comparisons with theoretical calculations and Monte Carlo simulations are also provided. In addition, the \LcDzero results are compared to $\Lambda/\kzero$ measurements in similar multiplicity intervals~\cite{Acharya:2019kyh}. The \pt-integrated \LcDzero yield ratios, extrapolated down to $\pt=0$, are also presented. 

\section{Experimental apparatus and data samples}

The ALICE experiment and its performance are presented in detail in Refs.~\cite{Aamodt:2008zz,Abelev:2014ffa}. The main detectors considered for the measurements discussed in this paper are the Inner Tracking System (ITS) for tracking, vertex reconstruction, event multiplicity estimation, and trigger purposes; the Time Projection Chamber (TPC) for tracking and particle identification; the Time-Of-Flight (TOF) for particle identification; and the V0 detector for event multiplicity estimation as well as for trigger purposes.

The event multiplicity selection was based on two estimators. At midrapidity ($|\eta|<1$) the multiplicity was estimated via the number of tracklets (\ntrkl) defined as track segments built by associating pairs of hits in the two Silicon Pixel Detector (SPD) layers, which are the two innermost layers of the ITS. The acceptance of the SPD in pseudorapidity changes with the longitudinal position of the vertex \zvtx and, in addition, the acceptance-times-efficiency changes with time due to variations of the inactive channels. Therefore, a data-driven correction procedure was applied on an event-by-event basis to \ntrkl, depending on the \zvtx position and the data taking period, as further described in Ref.~\cite{Adam:2015ota}. The event multiplicity in the forward rapidity region was estimated from the percentile distribution \vzeromperc of the V0M amplitude, which is the sum of signal amplitudes in the V0A and V0C scintillators. They are the two detecting components of the V0 detector on opposite sides of the interaction point along the beam axis, covering the pseudorapidity regions of $2.8<\eta<5.1$ and $-3.7<\eta<-1.7$, respectively. The \vzeromperc values towards 0 correspond to the highest multiplicity events,  while the lowest are assigned a value towards 100\%.

The data from \pp collisions at \thirteen used for this analysis were collected in the years 2016, 2017, and 2018. Three trigger setups were employed. The minimum-bias (MB) trigger required signals in both V0A and V0C in coincidence with the proton bunch arrival time. To enrich the data sample in the highest multiplicity regions, high-multiplicity triggers based on a minimum selection of the number of hits in the SPD (HMSPD) or of V0 amplitudes (HMV0) were used, which were fully efficient for $\ntrkl > 65$ and $\vzeromperc<0.1$\%, respectively.

Offline selection criteria were applied in order to remove background events from beam–gas collisions and other machine-induced background as described in Ref.~\cite{Acharya:2020kyh_2}. To reduce the contamination from events with superposition of more than one collision within the colliding bunches (pile-up), events with multiple reconstructed primary vertices were rejected. The impact of potentially remaining pile-up events is on the percent level and does not influence the final results of the present analysis. Only events with a vertex position of $|\zvtx|<10$~cm around the nominal interaction point were considered to ensure a uniform acceptance. In addition, events were required to have at least one reconstructed tracklet within the pseudorapidity region $|\eta|<1$ (\inel event class). This class of events minimises diffractive corrections and has a high trigger efficiency. It corresponds to about $75$\% of the total inelastic cross section~\cite{Acharya:2020kyh_2, Adam:2015gka}. After the aforementioned selections, the integrated luminosity of the data sample is about 32~$\rm{nb^{-1}}$ for the MB triggered events. Only the data periods granting an uniform efficiency of the HMV0 and HMSPD triggers, inside the range covered by the multiplicity intervals considered in the analysis, were used, resulting in an integrated luminosity of about 7.7~$\rm{pb^{-1}}$ with HMV0 and 0.8~$\rm{pb^{-1}}$ for the HMSPD trigger sample.

The events were assigned to multiplicity intervals based on the corresponding observables \ntrkl and \vzeromperc, as presented in Table~\ref{tab:multbins}. The last \ntrkl and \vzeromperc intervals contain data collected with the HMSPD and HMV0 triggers, respectively. To account for a possible trigger inefficiency for HMSPD triggered events in the range $60<\ntrkl<65$, a correction was applied with a data-driven reweighting procedure, as described in Ref.~\cite{Adam:2015ota}.

The mean multiplicity density (\avdndetaetarange) of charged primary particles, whose definition is given in Ref.~\cite{ALICE-PUBLIC-2017-005}, was obtained by converting the measured event multiplicities as described in Ref.~\cite{Acharya:2020kyh_2}. For the \vzeromperc percentiles the values reported in Ref.~\cite{Acharya:2020kyh_2} were used. The conversion of the specific \ntrkl intervals used in this analysis was performed by means of a PYTHIA~\cite{Sjostrand:2014zea} Monte Carlo (MC) simulation, with particle transport based on the GEANT3 package~\cite{Brun:1994aa}, and by selecting the charged primary particles measured at midrapidity in the events corresponding to the given \ntrkl intervals. Throughout the analysis reported in this paper, PYTHIA~8.243 with Monash tune~\cite{Skands:2014pea} was used; the version will not be reported later for the sake of simplicity.

A summary of the above information is given in Table~\ref{tab:multbins} together with the trigger correction $\epsilon^\mathrm{INEL}$ to account for those events which fulfil the \inel requirement but were not selected by the trigger, as specified in Ref.~\cite{Acharya:2020kyh_2}.

\renewcommand\arraystretch{1.3}
\begin{table}[tb!]
  \centering
  \caption{Summary of the multiplicity event classes at midrapidity (\ntrkl) and forward rapidity (\vzeromperc~[\%]), the latter corresponding to the visible V0M cross section. The average charged-particle densities \avdndetaetarange at midrapidity are shown, together with the value corresponding to the \inel event class. The trigger efficiency $\epsilon^\mathrm{INEL}$ is also reported for each multiplicity interval, as estimated in Ref.~\cite{Acharya:2020kyh_2}.}
  \begin{tabular}{|c|c|c|c|}
    \hline
    Mult. estimator & Mult. interval & \avdndetaetarange & $\epsilon^\mathrm{INEL}$ \\
    \hline
  
    \multirow{4}{*}{\ntrkl} & $[1,9]$ & $3.10\pm 0.02$ & $0.862\pm 0.015$\\
                            & $[10,29]$ & $10.54\pm 0.01$  & $0.997 \pm 0.002$\\
                            & $[30,59]$ & $22.56\pm 0.07$  & 1 (negl. unc.)\\
                            & $[60,99]$ & $37.83\pm 0.06$ & 1 (negl. unc.) \\
    \hline
    \multirow{3}{*}{\vzeromperc [\%]} & $[30,100]$ & $4.41\pm 0.05$ & $0.897 \pm 0.013$ \\
                                      & $[0.1,30]$ & $13.81\pm 0.14$ & $0.997 \pm 0.001$\\
                                      & $[0,0.1]$ & $31.53\pm 0.38$ & 1 (negl. unc.)  \\
    \hline
    \multicolumn{2}{|c|}{ \inel} & $6.93\pm 0.09$ & $0.920 \pm 0.003$ \\
    \hline
  \end{tabular}
  \label{tab:multbins}
\end{table}

\section{Data Analysis} 

The \Dzero, \Ds, and \Lc hadrons and their charge conjugates were reconstructed via the hadronic decay channels \DtoKpi (branching ratio ${\rm BR} = (3.950 \pm 0.031) \% $), \Dstophip (${\rm BR} = (2.24 \pm 0.08) \% $), \LctopKpi(${\rm BR} = (6.28 \pm 0.32) \% $), and $\LctopKzeros \to p\pi^+\pi^-$ (${\rm BR} = (1.10 \pm 0.06) \% $)~\cite{Zyla:2020zbs}. The analysis was performed for the different multiplicity intervals, as defined in Table~\ref{tab:multbins}. Transverse-momentum intervals between 1 and 24~\gevc were chosen to guarantee a large statistical significance in all multiplicity event classes. In order to minimise systematic effects, which could have a different impact in the different multiplicity intervals considered in the analysis, the same event and candidate selection criteria were used in all the multiplicity classes.
The charm-hadron decay tracks were excluded from the \ntrkl estimation at midrapidity, in order to reduce the effects of auto-correlation that could arise from the measurement of the charged-particle distribution in the same pseudorapidity region as the charm hadrons.
A possible remaining bias 
could be induced by the charged particles produced in the fragmentation of the charm quarks or by decays of excited charm states that are not subtracted from the \ntrkl count.

Candidates of \DtoKpi, \Dstophip, and \LctopKpi were defined by combining pairs or triplets of tracks with the proper charge signs, while the reconstruction of the \LctopKzeros candidates relied on reconstructing the V-shaped decay of the \kzero meson into two pions, which was then combined with a proton-candidate track. Track-quality selections were applied to the candidate daughters as explained in Ref.~\cite{PhysRevC.104.054905}. As a consequence of these track-selection criteria, the detector acceptance for D mesons and \Lc baryons varies as a function of rapidity, falling steeply to zero for $|y| > 0.5$ at low \pt and for $|y| > 0.8$ at $\pt > 5$~\gevc. For this reason, a fiducial acceptance selection was applied on the rapidity of the candidates, $|y| < y_{\mathrm{fid}}(\pt)$, where the factor $y_{\mathrm {fid}}(\pt)$ was defined as a second-order polynomial function, increasing from 0.5 to 0.8 in the transverse-momentum range $0< \pt < 5$~\gevc, and as a constant term, $y_{\mathrm{fid}}=0.8$, for $\pt > 5$~\gevc. The correction factors for the acceptance were computed accordingly. Further selections on the charm-hadron decay topology and on the particle identification (PID) of their decay products were exploited to reduce the combinatorial background. The same selection criteria described in Refs.~\cite{Acharya:2019mgn,PhysRevC.104.054905} were used for \Dzero and \LctopKpi, while for the \Ds and \LctopKzeros analyses, a machine-learning approach with Boosted Decision Trees (BDTs), using the toolkit from XGBoost~\cite{Chen:2016:XST:2939672.2939785}, was employed. Binary BDT classifiers were used and the training sample was assembled considering the background from the sidebands of the candidate invariant-mass distribution in data, and the prompt signal candidates from MC simulations based on the PYTHIA Monash event generator. Independent BDTs were trained for each \pt interval in the multiplicity-integrated sample.
The most prominent variables that were used in the training for the \Lc analysis are related to the PID of the proton decay track, the reconstructed invariant mass and $c\tau$ of the \kzero candidate, the cosine of the pointing angle between the line of flight of the \kzero meson (the vector connecting the primary and secondary vertices) and its reconstructed momentum vector, and the distance between the \kzero-meson decay vertex and the primary vertex. For the \Ds analysis, the variables provided to the BDTs are the same as reported in Ref.~\cite{Acharya:2021cqv}.
The selections on the BDT outputs were tuned to provide a large statistical significance for the signal.

The signal extraction was performed via binned maximum-likelihood fits to the invariant-mass distributions of candidates in each \pt and multiplicity interval. For all analyses, a Gaussian function was used to describe the signal peak. To model the background, an exponential function was used for the \Dzero mesons and for \Ds mesons with a transverse momentum higher than 4~\gevc, while a second-order polynomial function was used for both \Lc decay channels as well for the lowest two \pt intervals of the \Ds-meson analysis. Due to the limited number of candidates in some multiplicity classes and the large combinatorial background, it was not possible to extract the raw yield in the full \pt range for all the multiplicity intervals: the range $1 < \pt < 2$~\gevc in the low and high multiplicity classes and $12 < \pt < 24$~\gevc in the low multiplicity class are missing, respectively, for the \Ds and \Lc analyses. Examples of the invariant-mass distributions for \Dzero, \Ds, \LctopKpi, and \LctopKzeros candidates for the different \pt and multiplicity intervals are reported in Ref.~\cite{publicnote}.

The corrected per-event yields were computed for each \pt and multiplicity interval as
\begin{equation}
  \label{eq:CorrYield}
  \frac{1}{N^{\rm ev}_{\rm mult}}\frac{{\rm d^2} N^{\rm hadron}_{\rm mult}}{{\rm d}y{\rm d}\pt}=
  \frac{\epsilon_{\rm mult}^{\rm INEL}}{N^{\rm ev}_{\rm mult}} \frac{1}{c_{\Delta y}(\pt) \times \Delta \pt}\frac{1}{{\rm BR}}\frac{\left.f_{\rm prompt}(\pt) \times \frac{1}{2}\times N^{\rm hadron,raw}_{\rm mult}(\pt)\right|_{|y|<y_{\rm fid}(\pt)}}{({\rm Acc}\times\epsilon)_{\rm prompt,mult}(\pt)},
\end{equation}
where $N^{\rm hadron,raw}_{\rm mult}$ is the raw yield (sum of particles and antiparticles) extracted in a given \pt and multiplicity interval. It is multiplied by the prompt fraction $f_{\mathrm{prompt}}$ in order to correct for the corresponding beauty-hadron decay contribution, and divided by the multiplicity-dependent prompt acceptance-times-efficiency, $({\rm Acc}\times\epsilon)_{\rm prompt, mult}$. It is further divided by a factor of two to obtain the charge-averaged yield, by the BR of the decay channel, the \pt-interval width ($\Delta\pt$), and the correction factor for the rapidity coverage $c_{\Delta y}$, computed as the ratio between the generated heavy-flavour hadron yield in $\Delta y = 2 y_{\mathrm{fid}}$ and that in $|y| < 0.5$. The factor $N^{\rm ev}_{\mathrm{mult}}$ denotes the number of recorded events in each multiplicity class, which is then corrected for the fraction of \INEL events that were not selected by the trigger, $\epsilon_{\rm mult}^{\rm INEL}$, whose values are reported in Table~\ref{tab:multbins}.

The geometrical acceptance of the detector times the reconstruction efficiency $({\rm Acc}\times\epsilon)$ includes the tracking, the PID, and the topological selection efficiencies, and it was obtained separately for prompt and feed-down hadrons. It was determined from pp collisions simulated with PYTHIA with Monash tune, with particle transport based on the GEANT3 package~\cite{Brun:1994aa}. To account for the multiplicity dependence of the efficiency, which is driven by the primary-vertex resolution improving with increasing multiplicity, the generated events were weighted based on the number of tracklets in order to match the distribution observed in data. The generated \Lc \pt spectrum used to calculate the efficiencies was weighted to reproduce the shape obtained from the PYTHIA CR-BLC tune, which describes the measured spectra better than the Monash tune as observed in Ref.~\cite{PhysRevC.104.054905}.

The estimated $({\rm Acc}\times\epsilon)_{\rm prompt,mult}$ varies between 0.5\% and 60\% depending on \pt and species, and increases with multiplicity~\cite{publicnote}. The largest difference with respect to the efficiency computed in the \INEL class is observed in $1 < \pt < 2$~\gevc, where it reaches 30\% for \Ds, while it steeply decreases to few percents with increasing \pt.

The $f_{\mathrm{prompt}}$ fraction was estimated as reported in Refs.~\cite{PhysRevC.104.054905,Acharya:2021cqv}, using (i) the beauty-quark production cross section from FONLL calculations~\cite{Cacciari:1998it, Cacciari:2012ny}, (ii) the $({\rm Acc}\times\epsilon)$ for feed-down charm hadrons, (iii) beauty-quark fragmentation fractions determined from LHCb data~\cite{Aaij:2019pqz} for $\mathrm{b}\to\Lb$ and from \ee measurements~\cite{Gladilin:2014tba} for $\mathrm{b}\to \mathrm{B}$, and (iv) modelling the decay kinematics with PYTHIA simulations. The $f_{\mathrm{prompt}}$ fraction was assumed to be independent of the event multiplicity and therefore computed for the minimum-bias event class. This assumption is justified by the expected weak dependence of the feed-down fraction with multiplicity~\cite{Adam:2015ota}, predicted also by PYTHIA, and the small variations of the efficiency for the feed-down component of charm hadrons observed in the simulation for the different multiplicity intervals. The values of $f_{\mathrm{prompt}}$ range from 0.81 to 0.97 depending on \pt and particle species.

\section{Systematic uncertainty evaluation}

Sources of systematic uncertainty on the measured corrected yields were studied following procedures similar to those described in detail in Refs.~\cite{PhysRevC.104.054905,Acharya:2019mgn} for the minimum-bias \Lambdac and D-meson analyses. The multiplicity-independent sources, i.e.~those related to the tracking efficiency, the PID selection and the simulated charm-hadron \pt spectra, are discussed first, and then those related to the multiplicity dependence of the analyses are addressed. 

The systematic uncertainties on the track-reconstruction efficiency depend on the candidate \pt and number of decay tracks of the candidate, and range from 3\% to 5\% for the \Dzero, and from 4\% to 8\% for the \Ds and \Lambdac.
The contribution due to the PID was investigated by varying the selection criteria. For the \Dzero and the \LctopKpi analyses, the studies were performed as described in Refs.~\cite{Acharya:2019mgn} and~\cite{PhysRevC.104.054905}, respectively, resulting in a negligible uncertainty for the \Dzero, and a 5\% uncertainty for the \LctopKpi. In the \Ds and \LctopKzeros analyses, where topological and PID selection variables are used simultaneously in the BDT, the uncertainties coming from the two sources are treated in a combined procedure as described further below.

The possible differences between the real and simulated charm-hadron \pt spectra result in a further source of systematic uncertainty. It was evaluated by reweighting the \pt shape from PYTHIA Monash for the \Dzero and \Ds analyses and from PYTHIA CR-BLC for the \Lambdac analyses to match the one from D-meson FONLL calculations~\cite{Cacciari:1998it, Cacciari:2012ny}. 
This contribution ranges from 1\% to 6\% for $\pt < 4$~\gevc, while it is negligible at higher \pt.

The selection efficiencies of the various hadron candidates rely on the description of the detector resolution and alignment in the simulation. Systematic effects arising from imperfections in the simulation are studied by repeating the \Dzero and \LctopKpi analyses using different selection criteria on the displaced decay topology. In the \Ds and \LctopKzeros analyses, the selections on the BDT outputs were varied instead, covering both the PID and the decay-topology selection efficiency. For both approaches, the variations are performed separately for the different multiplicity and \pt intervals. The assigned systematic uncertainties are larger at low \pt where the selection criteria are strict, reaching 5\% for the \Dzero meson and 10\% for the \Ds and \Lambdac analyses.
The uncertainty due to the multiplicity dependence of the selection efficiency was evaluated as well, by changing the weight functions used to reproduce the measured charged-particle multiplicity in the simulation~\cite{Adam:2016mkz}. A maximum deviation of about 4\% is observed at low \pt and low multiplicity.

The systematic uncertainty on the raw-yield extraction was evaluated in each combination of the studied \pt and multiplicity intervals by repeating the fit to the invariant-mass distributions varying the fit range and the background fit function as done in Ref.~\cite{Acharya:2019mgn}. In order to test the sensitivity to the functional form of the fit function of the signal, the same strategy was performed using a bin-counting method, in which the signal yield was obtained from integrating the background-subtracted invariant-mass distribution. This systematic uncertainty ranges between 2\% and 14\% depending on the hadron species, the \pt, and the multiplicity interval.
 
As described above, a data-driven event reweighting procedure was applied for the HMSPD triggered data sample to account for the trigger inefficiency. Three strategies were explored to ensure normalised weights as outlined in Ref.~\cite{Adam:2015ota}. The different normalisation procedures were propagated to the raw yield calculation resulting in a relative systematic difference of 1\% to 4\% compared to the central values depending on the particle species, independent of their \pt.
  
Possible differences between the primary-vertex position distributions along the beam axis, \zvtx, in simulations and in data were investigated, since a slight dependence of the efficiencies with \zvtx is observed.
Hence, a further data-driven reweighting procedure was performed, taking this effect into account. A \pt-dependent systematic uncertainty was estimated, resulting in a contribution of about 0.5\% for $\pt<4$~\gevc, and negligible elsewhere. This systematic source is considered particle dependent because the weights are defined by selecting events with a charm-hadron candidate in a given invariant-mass range, for each hadron independently.

Systematic effects due to the dependence of the efficiency on the \ntrkl interval limits were also studied. These effects were a consequence of removing the reconstructed candidate’s decay tracks from the multiplicity in data but not in MC, which was done as the efficiencies have little dependence on multiplicity. The systematic uncertainty was evaluated by comparing the efficiency computed in a \ntrkl interval shifted by one or two units (for two- or three-body decays, respectively) with the one in the default intervals. It was observed to range from 2\% to 8\%, especially affecting the lowest \pt and multiplicity interval.

Two systematic uncertainties were assigned to the prompt fraction calculations, coming from the FONLL calculations of the b-quark production~\cite{Cacciari:1998it, Cacciari:2012ny} and the assumed multiplicity independence of the $f_{\rm prompt}$ factor. The FONLL parameters (b-quark mass, factorisation, and renormalisation scales) were varied as prescribed in Ref.~\cite{Cacciari:2012ny}. The assigned uncertainty values for the D mesons range from 3\% to 12\%. In the \Lambdac analyses, the additional contribution from the $f_{\rm b\rightarrow \Lambdab}$ fragmentation fraction is considered, as discussed in detail in Ref.~\cite{PhysRevC.104.054905}. It leads to more asymmetrical values of the uncertainty, ranging from $^{+2}_{-4}\%$ at low \pt to $^{+6}_{-8}\%$ at high \pt.
As mentioned above, Eq.~\ref{eq:CorrYield} describes the corrected prompt yields under the assumption that $f_{\rm prompt}$ does not vary with multiplicity. To estimate the uncertainty related to this assumption, PYTHIA simulations where employed, with Monash and CR-BLC tunes. The feed-down contribution from beauty-hadron decays, $f_{\rm feed{\text-}down} = 1 - f_{\rm prompt}$, was varied in each multiplicity interval based on the observed $f_{\rm feed{\text-}down}^{\rm mult} / \langle f_{\rm feed{\text-}down} \rangle$ trends in simulations. The feed-down contributions were found to be compatible for the D and \Lambdac hadrons and show a global increasing trend from 0.7 to 1.5 from the lowest to the highest multiplicity event class. The resulting systematic uncertainties depend on the charm-hadron species, the \pt interval, and the multiplicity classes considered in the analyses. For the part related to the $f_{\rm prompt}$ multiplicity-dependence assumption, typical values for the uncertainty for intermediate \pt are $^{+8}_{-0}\%$ at low multiplicity and $^{+0}_{-15}\%$ at high multiplicity.

The statistical uncertainty on the selection efficiency is assigned as systematic uncertainty.
It strongly depends on the \pt and multiplicity intervals, especially affecting the $\pt < 4$~\gevc and highest multiplicity intervals, where it reaches 1\% for the \Dzero, 4\% for the \Ds and \LctopKpi, and 5\% for the \LctopKzeros analysis.
Finally, an overall normalisation systematic uncertainty induced by the branching ratios~\cite{Zyla:2020zbs} was considered.

The sources of systematic uncertainty described above are assumed to be uncorrelated among each other and the total systematic uncertainty in each \pt and multiplicity interval is calculated as the quadratic sum of the estimated values. Depending on the \pt and multiplicity intervals, the resulting values range from 7\% to 13\% for the \Dzero, from 10\% to 17\% for the \Ds, from 7\% to 24\% for the \LctopKpi, and from 8\% to 17\% for the \LctopKzeros analyses.

\section{Results}

The \pt-differential corrected yield of the \Lambdac baryon was obtained in the different event-multiplicity classes, averaging the results from the two decay channels \LctopKpi and \LctopKzeros to obtain a more precise measurement, for which the inverse of the quadratic sum of the relative statistical and uncorrelated systematic uncertainties were used as weights. In the propagation of the uncertainties, the correlation between the statistical and systematic uncertainties was taken into account, with the strategy explained in Ref.~\cite{PhysRevC.104.054905}. In addition, the multiplicity-dependent systematic sources were considered as correlated between the two decay channels. In the rest of this section, \Lc will refer to the weighted average of the \LctopKpi and \LctopKzeros decay channels.

\begin{figure}[tb!]
  \begin{center}
    \includegraphics[width=1.0\textwidth]{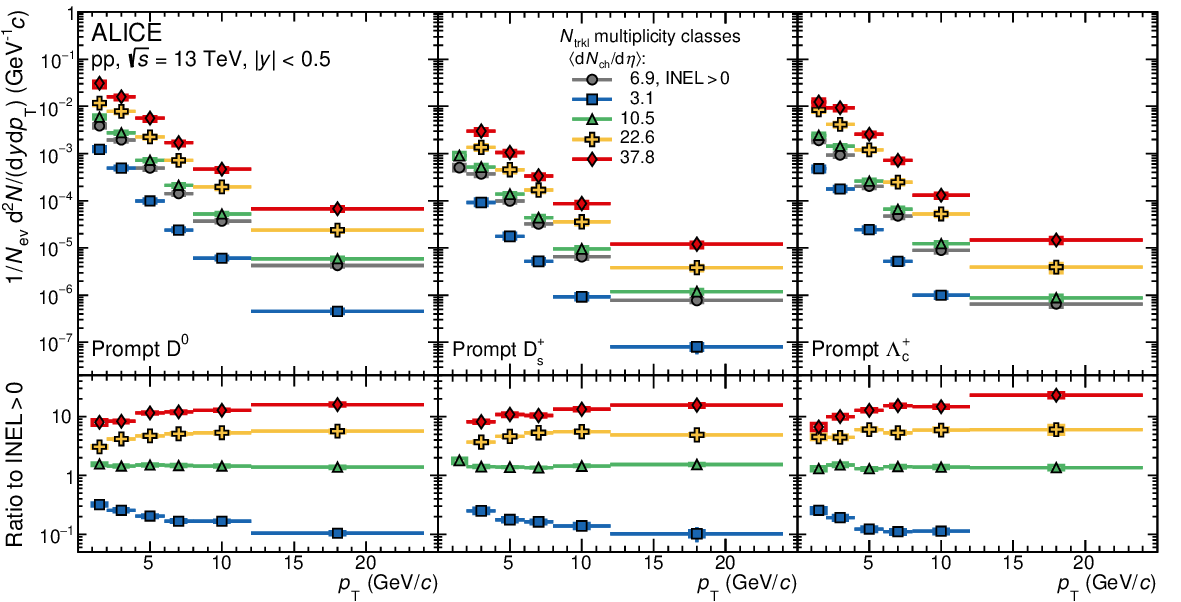}
    \caption{Transverse-momentum spectra of \Dzero, \Ds, and \Lc hadrons measured in \pp collisions at \thirteen for different multiplicity classes selected with the \ntrkl estimator at midrapidity. The corresponding ratios to \inel are shown in the bottom panels.}
    \label{fig:corryield_SPD}
  \end{center}
\end{figure}

The \pt-differential spectra of \Dzero, \Ds, and \Lambdac hadrons, measured in $|y|<0.5$, are shown in Fig.~\ref{fig:corryield_SPD} for the \inel class and the four multiplicity classes selected using the \ntrkl estimator at midrapidity. The statistical and total systematic uncertainties are shown by vertical error bars and boxes, respectively, as for all the figures in this section. The \pt spectra of the individual decay channels \LctopKpi and \LctopKzeros, as well as the \Dzero, \Ds, and \Lambdac yields in the multiplicity classes selected using the \vzeromperc estimator at forward rapidity, are reported in Ref.~\cite{publicnote}. The bottom panels of Fig.~\ref{fig:corryield_SPD} present the ratios to the \inel class, for which the multiplicity-dependent systematic sources were considered as uncorrelated among different multiplicity classes and the contributions of the tracking and PID efficiency, the shape of MC \pt spectra and \zvtx distribution, the beauty feed-down, and the branching ratio as correlated. The statistical uncertainties and the systematic uncertainties related to the selection efficiency and to the raw-yield extraction were considered partially correlated with respect to the measurement performed in the \inel class.

The measured \pt-differential yields increase from the lowest to the highest multiplicity class for the three hadron species. Their ratios to \inel increase (decrease) with increasing \pt for the highest (lowest) multiplicity class, suggesting a plateau towards $\pt > 10$~\gevc, as recently observed for the light-flavour hadrons in Refs.~\cite{Acharya:2020zji, Acharya:2019kyh, Acharya:2018orn}, where it was explained as a hardening of the measured \pt spectra with increasing \dNchdeta. Different MPI models were able to describe such effects~\cite{ALICE:2019dfi}, and it was shown to be more pronounced for protons than for kaons and pions, while similar for strange baryons and mesons. 

\begin{figure}[tb!]
  \begin{center}
    \includegraphics[width=0.85\textwidth]{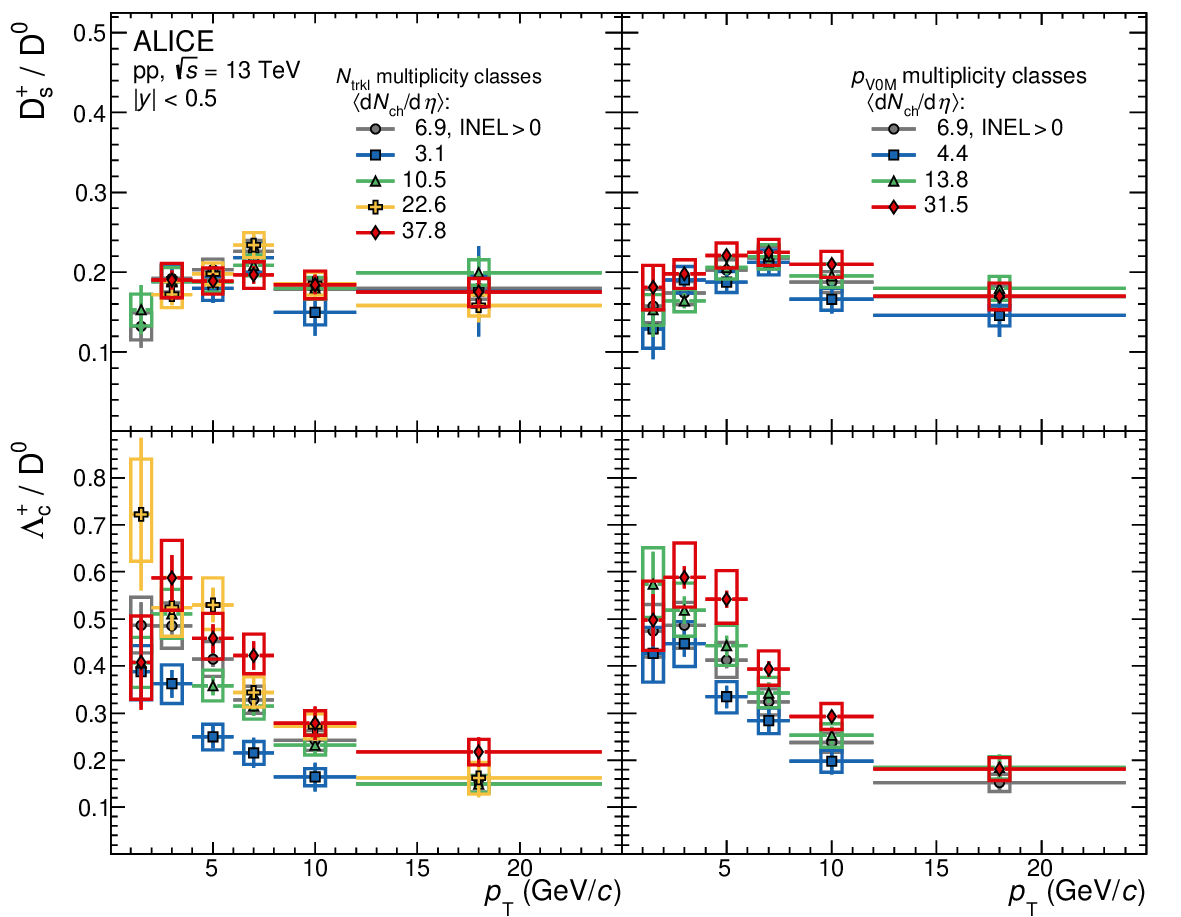}
    \caption{The \DsDzero (top) and \LcDzero (bottom) ratios measured in \pp collisions at \thirteen for different multiplicity classes at mid- (left) and forward (right) rapidity.}
    \label{fig:charmhadron_ratios}
  \end{center}
\end{figure}

In order to investigate potential differences in the \dNchdeta dependence of the \Dzero-meson production with respect to the \Ds meson and \Lambdac baryon, the \DsDzero and \LcDzero yield ratios are compared in different multiplicity event classes in Fig.~\ref{fig:charmhadron_ratios}, considering both forward and midrapidity multiplicity estimators. The sources of uncertainty assumed to be uncorrelated between different charm-hadron species included the raw-yield extraction, the selection efficiency, the shape of the MC \pt spectra, the \zvtx distribution, and the branching ratio. The systematic uncertainty deriving from the variation of the multiplicity-interval limits was propagated as partially correlated, while the other systematic uncertainties were assumed to be fully correlated. 

Within the current experimental uncertainties, the \DsDzero ratios are independent of \pt in the measured \pt range. They are compatible with measurements performed in pp collisions at $\sqrts=5.02$ and 7 TeV~\cite{Acharya:2019mgn}, and also with the average of the \pt-integrated results from experiments at \ee and \ep colliders, $0.17 \pm 0.03$~\cite{Gladilin:2014tba,Lisovyi:2015uqa}. A dependence of these ratios with multiplicity, as seen for the ratio of (multi-)strange hadrons to ${\rm \pi^{\pm}}$~\cite{ALICE:2017jyt,Acharya:2019kyh}, is not observed within the uncertainties. 

The \pt-differential \LcDzero ratios show an evident dependence on multiplicity, and a hierarchy is observed going from the lowest to the highest multiplicity intervals, for both the \ntrkl and \vzeromperc estimators, for all but the first \pt bin. The increasing trend with \dNchdeta for the \LcDzero ratio is consistent among the measurements done with the two multiplicity estimators, indicating that the enhancement between low and high multiplicity intervals is not a consequence of a possible bias arising from the coinciding rapidity regions between the multiplicity estimator and the measurement of interest at midrapidity. 
It is worth noticing that the measured \LcDzero ratio in the lowest multiplicity class is still higher, in the measured \pt range, than the average of corresponding ratios measured in \ee collisions at LEP, which was found to be $0.113\pm0.013{\rm(stat)}\pm0.006{\rm(syst)}$~\cite{PhysRevC.104.054905, Gladilin:2014tba}.

In order to estimate a significance level for the difference observed in the two extreme multiplicity classes at midrapidity, the highest multiplicity (HM) over the lowest multiplicity (LM) \LcDzero ratio was computed. The probability of the measured double-ratio ${\rm DR} = (\LcDzero)_{HM}/(\LcDzero)_{LM} > 1$, corresponds to a significance of 5.3$\sigma$ in the $1<\pt<12$~\gevc interval, considering as null hypothesis ${\rm DR} = 1$. This estimate was performed taking into account statistical and systematic uncertainties, for which the raw-yield extraction, the selection efficiency, the shape of the MC \pt spectra, and the \zvtx distribution sources were considered as uncorrelated, the multiplicity-interval limits as partially correlated, while the other sources cancelled out in the double ratio. With the aim of investigating the least favourable case, the measured values in all \pt intervals were shifted down by one standard deviation, by considering the sources of systematic uncertainties correlated with \pt that do not cancel out in the double ratio, i.e.~those arising from the selection efficiency and the generated \pt spectra.

\begin{figure}[tb!]
  \begin{center}
    \includegraphics[width=0.85\textwidth]{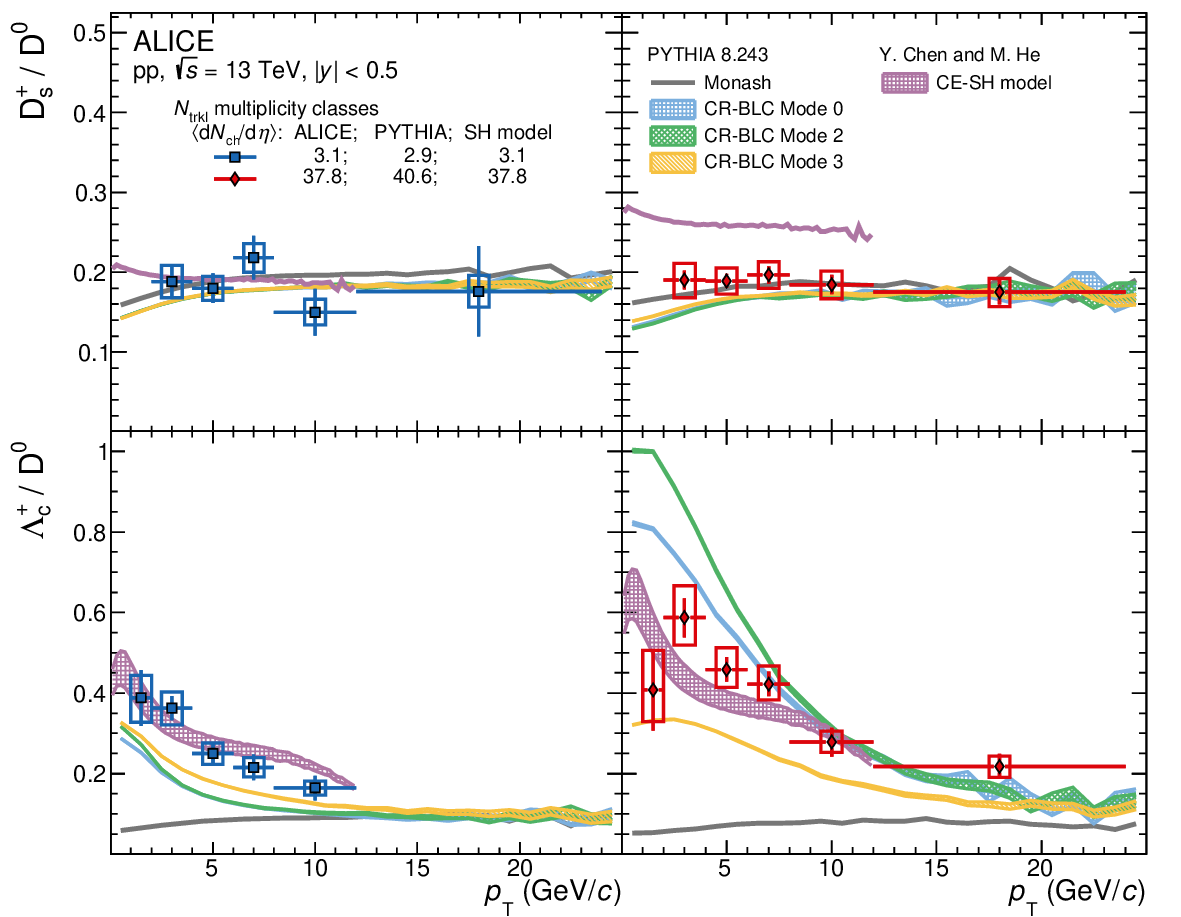}
    \caption{The \DsDzero (top) and \LcDzero (bottom) ratios measured in \pp collisions at \thirteen for the lowest (left) and highest (right) multiplicity classes at midrapidity. The measurements are compared to PYTHIA predictions with the Monash~\cite{Skands:2014pea} and the CR-BLC tunes~\cite{Christiansen:2015yqa}, and the CE-SH model~\cite{Chen:2020drg}, estimated in similar multiplicity classes. The uncertainty bands for the PYTHIA predictions are the statistical uncertainties on the simulations, while for the CE-SH model they refer to the variation of the branching ratios of the additional charm-baryon states from RQM~\cite{Ebert:2011kk}.}
    \label{fig:charmhadron_ratios_wtheory}
  \end{center}
\end{figure}

The measured charm-hadron ratios for the lowest and highest multiplicity class for the \ntrkl multiplicity estimator are compared to model predictions from MC generators and a statistical hadronisation model in Fig.~\ref{fig:charmhadron_ratios_wtheory}. The simulations with the PYTHIA event generator were performed with the Monash and the CR-BLC tunes. For the latter, three modes are suggested by the authors, applying different constraints on the allowed reconnections among colour sources, in particular concerning the causality connection among strings involved in a reconnection, and time dilation caused by relative boosts of the strings~\cite{Christiansen:2015yqa}. The simulations are shown in intervals of primary particle multiplicities selected at midrapidity, evaluated by studying the correlation between \ntrkl intervals and $N_{\rm ch}$ values. The estimated intervals are $1 \leq N_{\rm ch} \leq 12$ and $N_{\rm ch}>75$ for the lowest and highest multiplicity event classes, respectively. The measured \DsDzero ratios at low and high multiplicity are compatible with PYTHIA with Monash and CR-BLC tunes. The Monash tune, however, does not reproduce the \LcDzero ratio, and furthermore it does not show a multiplicity dependence. By contrast, the CR-BLC tunes describe the \LcDzero decreasing trend with \pt, and are closer to the overall magnitude, as also observed for minimum-bias \pp collisions at $\sqrts=5.02$ and 13 TeV~\cite{PhysRevC.104.054905,Acharya:2021vpo}. The CR-BLC tunes show a clear dependence with multiplicity, qualitatively reproducing the trend observed in data.

The measurements in Fig.~\ref{fig:charmhadron_ratios_wtheory} are also compared with the predictions of a canonical-ensemble statistical hadronisation (CE-SH) model~\cite{Chen:2020drg}, where the authors generalise the grand-canonical statistical hadronisation model (SHM)~\cite{He:2019tik} of charm-hadron production to the case of canonical SHM, and explore the multiplicity dependence of charm-hadron particle ratios. The version of the SHM model based on the measured charm-baryon spectrum reported by the PDG~\cite{Zyla:2020zbs} was observed to strongly underestimate the \LcDzero measurements in minimum-bias pp collisions~\cite{PhysRevC.104.054905}. For this reason, for the \LcDzero case, the underlying charm-baryon spectrum in the calculations is augmented to include additional excited baryon states predicted by the Relativistic Quark Model (RQM)~\cite{Ebert:2011kk}. For the \DsDzero predictions, only the PDG case is shown, since the RQM does not modify the D-meson yields with respect to the PDG set. The model calculations describe the \LcDzero ratios, reproducing the multiplicity dependence. The \DsDzero prediction is compatible with the measurement for the low multiplicity class, while it overestimates the data in the highest multiplicity interval. 
The CE-SH model explains the multiplicity dependence as deriving from the reduced volume size of the formalism towards smaller multiplicity, where a decrease of the \LcDzero ratio is a consequence of the strict baryon-number conservation. Such behaviour is also predicted for charm-strange mesons relative to charm mesons, based instead on strangeness-number conservation.

\begin{figure}[tb!]
  \begin{center}
    \includegraphics[width=1.0\textwidth]{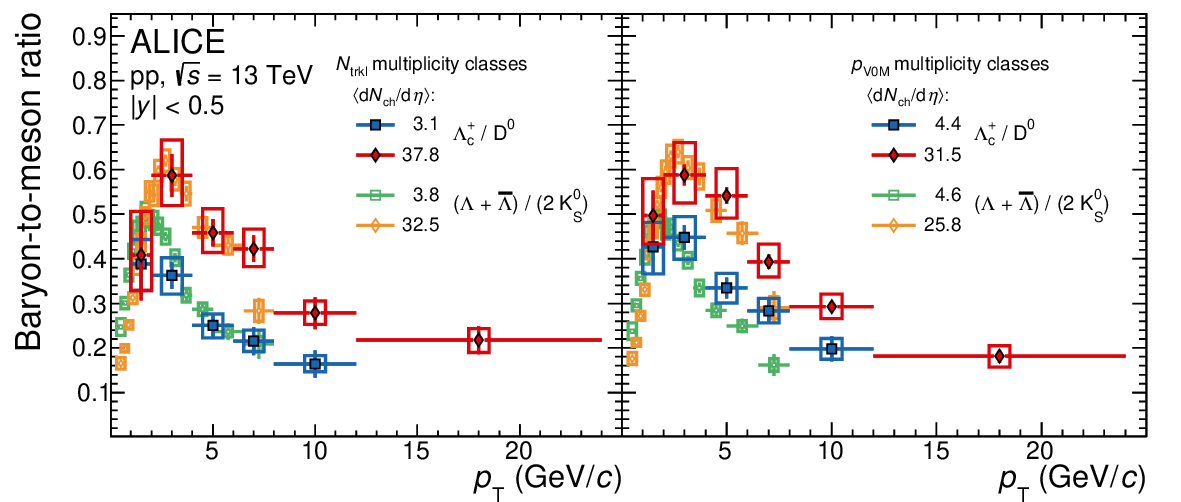}
    \caption{The baryon-to-meson ratios in the light-flavour, based on measurements from Ref.~\cite{Acharya:2019kyh} and charm sector measured in \pp collisions at \thirteen for similar low- and high-multiplicity classes at mid- (left) and forward (right) rapidity.}
    \label{fig:lcdzero_vs_lk0s}
  \end{center}
\end{figure}

Figure~\ref{fig:lcdzero_vs_lk0s} shows the comparison of the \LcDzero and the $\rm \Lambda/\kzero$~\cite{Acharya:2019kyh} baryon-to-meson ratios as a function of \pt in \pp collisions at \thirteen, in similar low and high \ntrkl and \vzeromperc multiplicity classes. In the vacuum-fragmentation scenario, the light-flavour hadron production has a significant contribution from gluon fragmentation, whereas heavy-flavour hadrons are primarily produced through the fragmentation of a charm quark, which is in turn produced in the initial hard scattering. In addition, at low \pt, light-flavour hadrons originate mainly from small-momentum soft scattering processes. Despite these differences, the light- and heavy-flavour baryon-to-meson ratios, \LcDzero and $\rm \Lambda/\kzero$, show a remarkably similar trend as a function of \dNchdeta. The measurements also suggest a similar shift of the baryon-to-meson ratio peaks towards higher momenta, with increasing multiplicity. These similarities, observed as well in minimum-bias \pp and \pPb collisions at \fivenn both in terms of shape and magnitude~\cite{PhysRevLett.127.202301}, hint at a potential common mechanism for light- and charm-baryon formation in hadronic collisions at LHC energies. 

\begin{figure}[tb!]
  \begin{center}
    \includegraphics[width=0.7\textwidth]{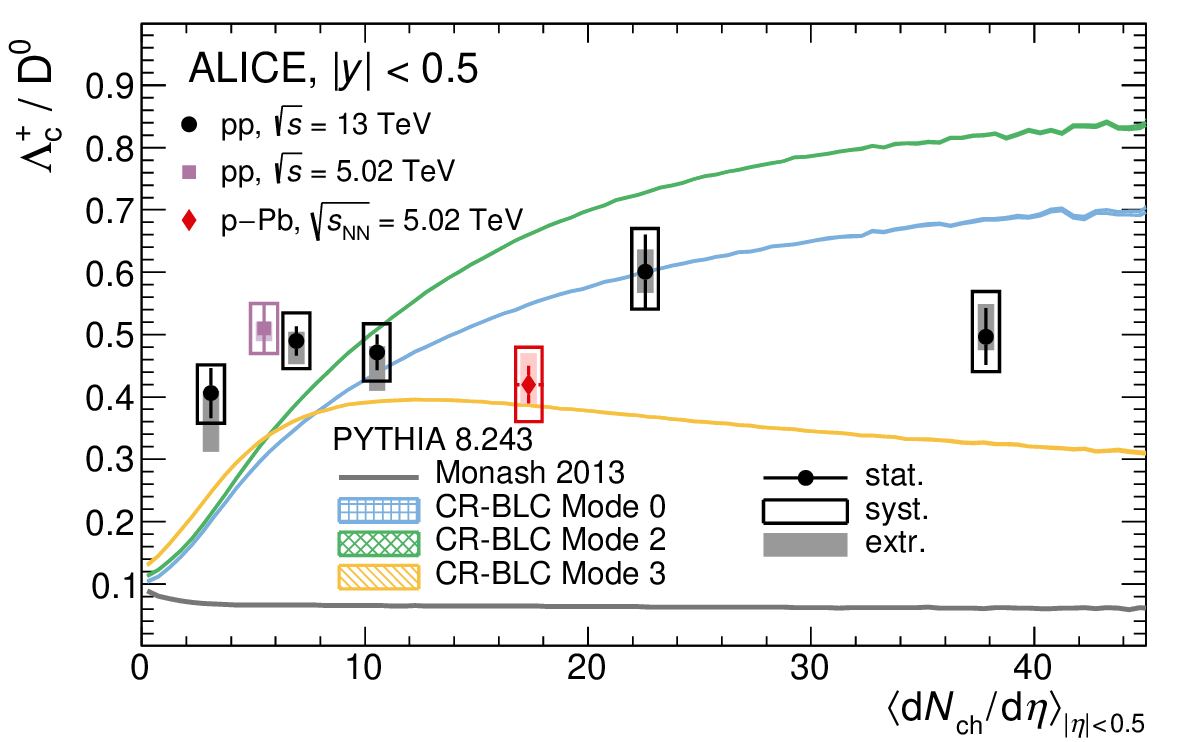}
    \caption{Ratios of \pt-integrated yields of \Lambdac and \Dzero hadrons as a function of \dNchdeta in pp collisions at \thirteen. 
    Measurements performed in pp and p--Pb collisions at \fivenn from Ref.~\cite{PhysRevC.104.054905} are also shown. Statistical and systematic uncertainties are shown by error bars and empty boxes, respectively. Shaded boxes represent the extrapolation uncertainties. The corresponding PYTHIA predictions~\cite{Skands:2014pea,Christiansen:2015yqa} are also shown.}
    \label{fig:ptint}
  \end{center}
\end{figure}

The \pt-integrated yields of \Lambdac and \Dzero were computed by integrating the \pt-differential spectra in their measured range and extrapolating them down to $\pt = 0$ in each multiplicity interval. In the integration, the systematic uncertainties were propagated considering the uncertainty due to the raw-yield extraction and the statistical uncertainty on the efficiency as fully uncorrelated and all the other sources as fully correlated among \pt intervals. The PYTHIA predictions with CR-BLC Mode~2 were used for the extrapolation in each multiplicity interval, for both \Lambdac and \Dzero, following a similar procedure as the one described in Ref.~\cite{PhysRevC.104.054905}. 
The extrapolation factor was computed as the ratio of the
PYTHIA spectrum integrated in the full \pt range to the integral in the visible \pt range. The \Lambdac and \Dzero yields in the full \pt range were obtained by integrating the yield in the visible \pt interval and scaling by the extrapolation factor. 
The fraction of extrapolated yield from the lowest to the highest multiplicity interval is about 39\% (31\%), 28\% (22\%), 20\% (16\%), and 15\% (13\%) for \Lambdac (\Dzero).
The procedure was repeated considering also the CR-BLC Mode~0 and Mode~3 as well as two different functions fitted to the spectra (a Tsallis-Lévy~\cite{Prato:1999jj} and a power-law function). The fits were performed considering the statistical and \pt-uncorrelated sources of systematic uncertainties, and also shifting up and down the data by one sigma of the \pt-correlated systematic uncertainties. The envelope of the extrapolation factors obtained with all the trials was assigned as the extrapolation uncertainty on \Lambdac and \Dzero, and it was propagated to the \LcDzero ratio, resulting in a value that ranges from 2\% to 21\% depending on multiplicity.
The same procedure was used to estimate the \pt-integrated \Ds yields and \DsDzero yield ratios in the different multiplicity intervals, reported in Ref.~\cite{publicnote}. 
The \Lambdac and \Dzero \pt-integrated yields are also reported in Ref.~\cite{publicnote}, together with the \pt-integrated \LcDzero yield ratios in the visible \pt range, and the tables with the numerical values of the \pt-integrated ratios. 
The \pt-integrated \LcDzero yield ratio as a function of \dNchdeta is shown in Fig.~\ref{fig:ptint}, where the systematic uncertainties from the extrapolation (shaded boxes, assumed to be uncorrelated among multiplicity intervals) are drawn separately from the other sources of systematic uncertainties (empty boxes). 
The sources related to the raw-yield extraction, the multiplicity-interval limits, the high-multiplicity triggers, the multiplicity-independent prompt fraction assumption, and the statistical uncertainties on the efficiencies are also considered uncorrelated with multiplicity. The other systematic uncertainties are assumed to be correlated.
The measurements performed in pp and p--Pb collisions at \five~\cite{PhysRevC.104.054905} are also shown. The result does not favour an increase of the yield ratios with multiplicity and the trend is compatible with a constant function. The same trend was also observed for the $\rm \Lambda/\kzero$ ratio in Ref.~\cite{Acharya:2019kyh}.  This suggests that the increasing trend observed for the $1<\pt<24$~\gevc range comes from a re-distribution of \pt that acts differently for baryons and mesons, while this is not observed in the meson-to-meson ratios, as shown in Fig.~\ref{fig:charmhadron_ratios_wtheory} for \DsDzero and in Ref.~\cite{Acharya:2018orn} for $\rm{K}/\rm{\pi}$. 
The results are compared to the \pt-integrated PYTHIA predictions. 
The measurements exclude the Monash prediction in the whole multiplicity range, and tend to be significantly below the CR-BLC Mode~2 for the three highest multiplicity intervals.

\section{Conclusions}

The first measurement of \DsDzero and \LcDzero ratios as a function of charged-particle multiplicity in pp collisions at \thirteen was presented. The \pt-differential \DsDzero yield ratio does not show a dependence on multiplicity, within uncertainties.  In contrast, the charm baryon-to-meson ratio, \LcDzero, measured as a function of \pt, shows a significant increase (5.3$\sigma$) when comparing the measurements performed in the lowest and highest multiplicity intervals in $1<\pt<12$~\gevc. In addition, the \LcDzero ratio measured in the lowest multiplicity interval ($\avdndeta=3.1$) is higher, at low and intermediate \pt, than the values measured at \ee colliders at lower centre-of-mass energies. These observations imply a modification of the hadronisation mechanisms that is collision-system and multiplicity dependent, further confirming the limited validity of the  assumption of universality of the fragmentation functions. The measurements are compared with two calculations. Those based on PYTHIA with CR-BLC describe the \DsDzero measurements and capture the trend of the \LcDzero ratio, qualitatively describing the increasing magnitude of the baryon-to-meson ratios with multiplicity.  Calculations based on a statistical hadronisation model, with the multiplicity dependence originating from the canonical treatment of quantum-charge conservation, describe the \LcDzero measurements in the lowest and highest multiplicity intervals. The prediction is also in agreement with the \DsDzero ratio for the low multiplicity interval, while it overestimates the data in the high-multiplicity class. The baryon-to-meson ratios in the charm sector, \LcDzero, are also compared to those in the light-flavour sector, $\rm \Lambda/\kzero$, in similar multiplicity classes, showing a remarkably similar trend as a function of \pt and similar enhancement with \dNchdeta. These similarities hint at a possible common mechanism for light- and charm-baryon formation in pp collisions at LHC energies. 
The \pt-integrated \LcDzero ratios, extrapolated to $\pt=0$ based on spectral shapes from PYTHIA with CR-BLC,  show no significant dependence on multiplicity, suggesting that the increase in the baryon-to-meson yield ratio observed in the measured \pt range is due to a different redistribution of \pt between baryons and mesons, rather than to an enhancement in the overall baryon yield. 
More precise measurements with the data sample collected in Run~3 of the LHC, that is planned to start late spring 2022, will allow us to further investigate the shape of the \pt-integrated baryon-to-meson ratios versus multiplicity, extending the multiplicity reach to lower and higher multiplicity intervals.


\newenvironment{acknowledgement}{\relax}{\relax}
\begin{acknowledgement}
\section*{Acknowledgements}

The ALICE Collaboration would like to thank all its engineers and technicians for their invaluable contributions to the construction of the experiment and the CERN accelerator teams for the outstanding performance of the LHC complex.
The ALICE Collaboration gratefully acknowledges the resources and support provided by all Grid centres and the Worldwide LHC Computing Grid (WLCG) collaboration.
The ALICE Collaboration acknowledges the following funding agencies for their support in building and running the ALICE detector:
A. I. Alikhanyan National Science Laboratory (Yerevan Physics Institute) Foundation (ANSL), State Committee of Science and World Federation of Scientists (WFS), Armenia;
Austrian Academy of Sciences, Austrian Science Fund (FWF): [M 2467-N36] and Nationalstiftung f\"{u}r Forschung, Technologie und Entwicklung, Austria;
Ministry of Communications and High Technologies, National Nuclear Research Center, Azerbaijan;
Conselho Nacional de Desenvolvimento Cient\'{\i}fico e Tecnol\'{o}gico (CNPq), Financiadora de Estudos e Projetos (Finep), Funda\c{c}\~{a}o de Amparo \`{a} Pesquisa do Estado de S\~{a}o Paulo (FAPESP) and Universidade Federal do Rio Grande do Sul (UFRGS), Brazil;
Ministry of Education of China (MOEC) , Ministry of Science \& Technology of China (MSTC) and National Natural Science Foundation of China (NSFC), China;
Ministry of Science and Education and Croatian Science Foundation, Croatia;
Centro de Aplicaciones Tecnol\'{o}gicas y Desarrollo Nuclear (CEADEN), Cubaenerg\'{\i}a, Cuba;
Ministry of Education, Youth and Sports of the Czech Republic, Czech Republic;
The Danish Council for Independent Research | Natural Sciences, the VILLUM FONDEN and Danish National Research Foundation (DNRF), Denmark;
Helsinki Institute of Physics (HIP), Finland;
Commissariat \`{a} l'Energie Atomique (CEA) and Institut National de Physique Nucl\'{e}aire et de Physique des Particules (IN2P3) and Centre National de la Recherche Scientifique (CNRS), France;
Bundesministerium f\"{u}r Bildung und Forschung (BMBF) and GSI Helmholtzzentrum f\"{u}r Schwerionenforschung GmbH, Germany;
General Secretariat for Research and Technology, Ministry of Education, Research and Religions, Greece;
National Research, Development and Innovation Office, Hungary;
Department of Atomic Energy Government of India (DAE), Department of Science and Technology, Government of India (DST), University Grants Commission, Government of India (UGC) and Council of Scientific and Industrial Research (CSIR), India;
Indonesian Institute of Science, Indonesia;
Istituto Nazionale di Fisica Nucleare (INFN), Italy;
Japanese Ministry of Education, Culture, Sports, Science and Technology (MEXT), Japan Society for the Promotion of Science (JSPS) KAKENHI and Japanese Ministry of Education, Culture, Sports, Science and Technology (MEXT)of Applied Science (IIST), Japan;
Consejo Nacional de Ciencia (CONACYT) y Tecnolog\'{i}a, through Fondo de Cooperaci\'{o}n Internacional en Ciencia y Tecnolog\'{i}a (FONCICYT) and Direcci\'{o}n General de Asuntos del Personal Academico (DGAPA), Mexico;
Nederlandse Organisatie voor Wetenschappelijk Onderzoek (NWO), Netherlands;
The Research Council of Norway, Norway;
Commission on Science and Technology for Sustainable Development in the South (COMSATS), Pakistan;
Pontificia Universidad Cat\'{o}lica del Per\'{u}, Peru;
Ministry of Education and Science, National Science Centre and WUT ID-UB, Poland;
Korea Institute of Science and Technology Information and National Research Foundation of Korea (NRF), Republic of Korea;
Ministry of Education and Scientific Research, Institute of Atomic Physics, Ministry of Research and Innovation and Institute of Atomic Physics and University Politehnica of Bucharest, Romania;
Joint Institute for Nuclear Research (JINR), Ministry of Education and Science of the Russian Federation, National Research Centre Kurchatov Institute, Russian Science Foundation and Russian Foundation for Basic Research, Russia;
Ministry of Education, Science, Research and Sport of the Slovak Republic, Slovakia;
National Research Foundation of South Africa, South Africa;
Swedish Research Council (VR) and Knut \& Alice Wallenberg Foundation (KAW), Sweden;
European Organization for Nuclear Research, Switzerland;
Suranaree University of Technology (SUT), National Science and Technology Development Agency (NSDTA) and Office of the Higher Education Commission under NRU project of Thailand, Thailand;
Turkish Energy, Nuclear and Mineral Research Agency (TENMAK), Turkey;
National Academy of  Sciences of Ukraine, Ukraine;
Science and Technology Facilities Council (STFC), United Kingdom;
National Science Foundation of the United States of America (NSF) and United States Department of Energy, Office of Nuclear Physics (DOE NP), United States of America.
\end{acknowledgement}

\bibliographystyle{utphys}   
\bibliography{bibliography}

\newpage
\appendix

%
%

\section{The ALICE Collaboration}
\label{app:collab}
\small
\begin{flushleft} 

\bigskip 

S.~Acharya$^{\rm 142}$, 
D.~Adamov\'{a}$^{\rm 96}$, 
A.~Adler$^{\rm 74}$, 
J.~Adolfsson$^{\rm 81}$, 
G.~Aglieri Rinella$^{\rm 34}$, 
M.~Agnello$^{\rm 30}$, 
N.~Agrawal$^{\rm 54}$, 
Z.~Ahammed$^{\rm 142}$, 
S.~Ahmad$^{\rm 16}$, 
S.U.~Ahn$^{\rm 76}$
% ...

\bigskip

\bigskip 

\textbf{\Large Affiliation Notes}

\bigskip 

% ...
$^{\rm VII}$ Also at: University of Kansas, Lawrence, Kansas, United States\\

\bigskip

\bigskip 

\textbf{\Large Collaboration Institutes}

\bigskip 

% ...
$^{16}$ Department of Physics, Aligarh Muslim University, Aligarh, India\\
% ...
$^{30}$ Dipartimento DISAT del Politecnico and Sezione INFN, Turin, Italy\\
% ...
$^{34}$ European Organization for Nuclear Research (CERN), Geneva, Switzerland\\
% ...
$^{54}$ INFN, Sezione di Bologna, Bologna, Italy\\
% ...
$^{74}$ Johann-Wolfgang-Goethe Universit\"{a}t Frankfurt Institut f\"{u}r Informatik, Fachbereich Informatik und Mathematik, Frankfurt, Germany\\
% ...
$^{76}$ Korea Institute of Science and Technology Information, Daejeon, Republic of Korea\\
% ...
$^{81}$ Lund University Department of Physics, Division of Particle Physics, Lund, Sweden\\
% ...
$^{96}$ Nuclear Physics Institute of the Czech Academy of Sciences, \v{R}e\v{z} u Prahy, Czech Republic\\
% ...
$^{142}$ Variable Energy Cyclotron Centre, Homi Bhabha National Institute, Kolkata, India\\
% ...

\bigskip 

\end{flushleft} 
  
\end{document}